\documentclass[aps,prl,floatfix]{revtex4}
\usepackage{graphicx,amsmath,units,xspace}

\graphicspath{{figures/}}

\begin{document}

\title{Phase-Space Networks of Geometrical Frustrated Systems}

\author{Yilong Han}
\affiliation{Department of Physics, Hong Kong University of Science
and Technology, Clear Water Bay, Kowloon, Hong Kong}
\date{\today}

\begin{abstract}
Geometric frustration leads to complex phases of matter with exotic
properties. Antiferromagnets on triangular lattices and square ice
are two simple models of geometrical frustration. We map their
highly degenerated ground-state phase spaces as discrete networks
such that network analysis tools can be introduced to phase-space
studies. The resulting phase spaces establish a novel class of
complex networks with Gaussian spectral densities. Although
phase-space networks are heterogeneously connected, the systems are
still ergodic except under periodic boundary conditions. We
elucidate the boundary effects by mapping the two models as stacks
of cubes and spheres in higher dimensions. Sphere stacking in
various containers, i.e. square ice under various boundary
conditions, reveals challenging combinatorial questions. This
network approach can be generalized to phase spaces of some other
complex systems.
\end{abstract}

\maketitle

\section{1. Introduction}

One challenge to understanding disordered solids is the complex
geometry of their phase spaces, including the relative positions and
interconnections between the different metastable states. Phase
spaces are usually too large and complicated to be directly studied.
For example, an $N$-particle system typically has a vast abstract
$6N$-dimension phase space ($3N$ for position, $3N$ for velocity).
Here, we propose that some simple models of disordered solids, such
as geometrical frustrated spin models, provide an ideal platform for
phase-space studies. Their phase spaces can be mapped as nontrivial
complex networks, so that the recently developed large tool box of
network analysis \cite{Albert02,Newman03,Costa07} can be used to
understand phase spaces. On the other hand, these phase spaces
provide a new class of complex networks with novel topologies.

When a system has competing interactions, there is no way to
simultaneously satisfy all interactions, a situation known as
frustration. Frustration widely exists in systems ranging from
neural networks to disordered solids. Frustration can also arise in
an ordered lattice solely from geometric incompatibility
\cite{Moessner06}. For example, consider the three antiferromagnetic
Ising spins on the triangle shown in Fig.~\ref{fig:cubenet}A. Once
two of them are antiparallel to satisfy their antiferromagnetic
interaction, there is no way that the third one can be antiparallel
to both of the other two spins. Frustration leads to highly
degenerated ground states and, subsequently, to complex materials
with peculiar dynamics such as water ice \cite{Pauling35}, spin ice
\cite{Bramwel01}, frustrated magnets \cite{Bramwel01}, artificial
frustrated systems \cite{Wang06} and soft frustrated materials
\cite{Han08}.

In geometrical frustrated systems, spins on lattices have discrete
degrees of freedom, such that their phase spaces are discrete and
can be viewed as networks. A node in the network corresponds to a
state of the system. Two nodes are connected by an edge (i.e. a
link) if the system can directly evolve from one state to the other
without passing through intermediate states. Edges are undirected
because dynamic processes at the microscopic level are time
reversible. The challenge is how to construct and analyze such large
phase-space networks. For example, how do we identify whether or not
two nodes are connected?

\section{2. Antiferromagnets on triangular lattices.}

\begin{figure}
\centering
\includegraphics[width=1\columnwidth]{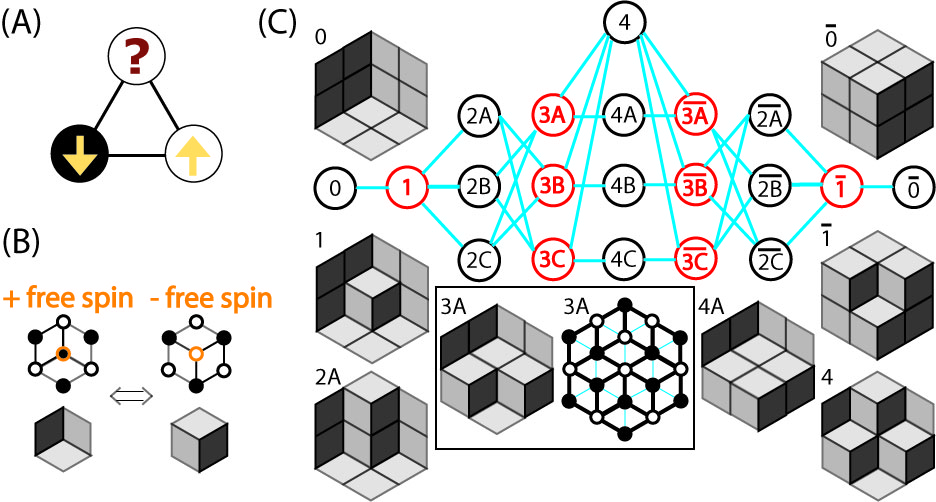}
\caption{\textbf{A phase-space network of cube stacks.} (A): Three
antiferromagnetic spins on a triangle cannot simultaneously satisfy
all their interactions. (B): The central spin has three up and three
down neighbours, so that it can flip freely without energy change.
Satisfied bonds can be viewed as cubes. The +/- free spin flip
corresponds to adding/removing a cube. (C): The $2 \times 2 \times
2$ cube stacks are stable against gravity along the [1,1,1]
direction. Stack configurations have one-to-one correspondence to
Ising ground states under `hexagon boundary condition', e.g., see
configuration 3A. In the right 3A configuration, the black lines are
satisfied bonds forming rhombuses and the blue lines are frustrated
bonds. In total, there are 20 legal stacks, i.e., 20 nodes in the
phase-space network. The network is bipartite, i.e., consisting of
alternating red (even number of cubes) and black (odd number of
cubes) states.}
 \label{fig:cubenet}
\end{figure}

The first model we consider is antiferromagnetic Ising spins on a
two-dimensional (2D) triangular lattice \cite{Wannier50}. For a
large system with periodic boundary conditions, it has $\sim
e^{0.323N_{spin}}$ degenerated ground states where $N_{spin}$ is the
number of spins \cite{Wannier50}. For example, configuration 3A in
Fig.~\ref{fig:cubenet}C is one ground state in the hexagonal area.
We refer to pairs of neighbouring spins in opposite states as
satisfied bonds, i.e., they satisfy the antiferromagnetic
interaction. Since one triangle has at most two satisfied bonds (see
Fig.~\ref{fig:cubenet}A), the ground state should have 1/3 of its
bonds frustrated and 2/3 of its bonds satisfied \cite{Wannier50}. If
we plot only satisfied bonds, a ground state can be mapped to a
random lozenge tiling \cite{Blote94}, see configuration 3A in
Fig.~\ref{fig:cubenet}C. A lozenge is a rhombus with $60^{\circ}$
angles. By colouring lozenges with different orientations with
different grey scales, the tiling can be viewed as a stack of 3D
cubes, or as a simple cubic crystal surface projected in the [1,1,1]
direction \cite{Blote94}, see Fig.~\ref{fig:cubenet}C.

The ground state has a local zero-energy mode, as shown in
Fig.~\ref{fig:cubenet}B: the central particle can flip without
changing the energy since it has 3 up and 3 down neighbours. The
system can evolve via a sequence of such single spin flips, even at
zero temperature. We call such a local zero-energy mode the
\textit{basic flip}. Any configuration change can be viewed as a
sequence of such basic flips. Recently, we directly observed such
flips in a colloidal monolayer \cite{Han08}. In the language of
cubes, a basic flip is equivalent to adding or removing a cube, see
Fig.~\ref{fig:cubenet}B. By continuing to add or remove one cube
from the stack surface, we can access all possible stack
configurations in the large box. Thus, the ground-state phase space
is connected by this `hexagonal boundary condition'. The
corresponding cube stacking in a large box is equivalent to the
boxed plane partition problem in combinatorics \cite{Andrews04}. The
total number of ways to stack unit cubes in an $L^3$ box is given by
the MacMahon formula: \cite{MacMahon}
\begin{eqnarray}
N_n(L)&=&\prod_{1\le i, j, k \le L}\frac{i+j+k-1}{i+j+k-2}
=\frac{H^3(L)H(3L)}{H^3(2L)} \nonumber \\
&\sim & \left(\frac{27}{16}\right)^{\frac{3}{2}L^2} \textrm{when }
L\to \infty, \label{eq:g}
\end{eqnarray}
where the hyperfactorial function $H(L)=\prod_{k=0}^{L-1} k!$. The
first several $N_n(L=2,3,4,5,\cdots)$ are 20, 980, 232848,
267227532,$\cdots$ (see the number sequence A008793 in
ref.~\cite{sequence}). When $L=2$, all 20 ground-state
configurations in Fig.~\ref{fig:cubenet}C have the same minimum
possible energy, i.e., 12 frustrated bonds in 12 rhombuses.

\section{3. Network properties.} The 20-node phase-space network shown
in Fig.~\ref{fig:cubenet}C can be constructed based on the following
two facts: (1) Spins have discrete degrees of freedom, such that the
phase space is a discrete network; (2) Any configuration change can
be decomposed to a sequence of basic flips. Consequently, we can
define an edge between two nodes if the two states differ by only
one basic flip (i.e., one cube), such that the system can
\textit{directly} change from one node to the other without passing
through intermediate nodes.

\begin{figure}
\centering
\includegraphics[width=1\columnwidth]{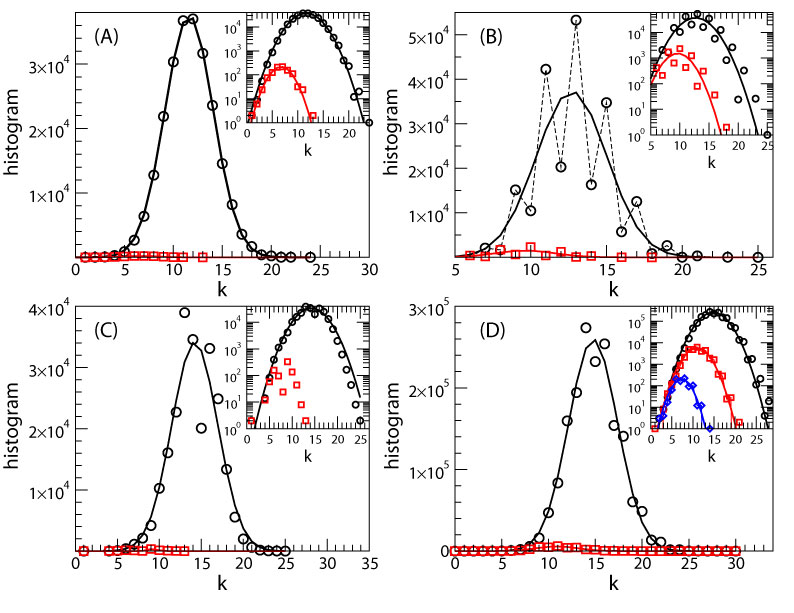}
\caption{\textbf{Connectivity distributions.} Histograms of the
connectivities of ground-state phase-space networks for (A)
antiferromagnets on triangular lattices, and (B, C, D) square ices
under different boundary conditions. (A): $L=4$ (circles) and $L=3$
(squares) cube stacks. (B): Spheres stacks in $L=6$ (circles) and
$L=5$ (squares) tetrahedra. (C): Spheres stacks in $L=4$ (circles)
and $L=3$ (squares) octahedra. (D): Sphere stacks in $L=6$
(circles), $L=5$ (squares) and $L=4$ (diamonds) containers shown in
Fig.~\ref{fig:alterboundary}. Insets: semi-log plots. The curves in
the main plots and insets show the best Gaussian fits.}
\label{fig:histok}
\end{figure}

Numerically, we can handle networks only up to $L=4$ stacks with
$N_n=232848$ nodes; nevertheless many general properties have
emerged from such small systems. Figure \ref{fig:histok}A shows the
connectivity (i.e. degree) distribution
\cite{Albert02,Newman03,Costa07} of cube-stack networks. The
connectivity, $k_i$, is the number of edges incident with the node
$i$. The connectivities of various frustrated systems appear to have
Gaussian distributions (see Fig.~\ref{fig:histok}). This behavior is
similar to that of small-world networks \cite{Watts98,Costa07} and
Poisson random networks \cite{Newman03,Costa07} and different from
that of scale-free networks \cite{Costa07,{Barabasi99}}.

Other network properties, such as the diameter and the cluster
coefficient, can be readily derived from the cube stack picture. The
shortest path length between two nodes is simply the number of
different sites among all the $L^3$ sites. The largest distance,
i.e. the diameter of the network, is $L^3$ between the `vacant' and
the `full' states. Here, we define the vacant state as no cube
(i.e., $L^3$ vacant sites) and the full state as no vacant site
(i.e., $L^3$ cubes). The networks have small-world properties
\cite{Watts98,Costa07} in the sense that the diameter, $L^3$, is
almost logarithmically small compared with the network size, $\sim
e^{N_{spin}}\sim e^{L^2}$. The network is bipartite (see the black
and red circles in Fig.~\ref{fig:cubenet}C) because a cube stack
comes back to its initial configuration only by adding and removing
the same number of cubes, i.e., an even number of basic flips.
Consequently, the cluster coefficient \cite{Costa07}, which
characterizes the density of triangles in the network, is 0.

\begin{figure}
\centering
\includegraphics[width=1\columnwidth]{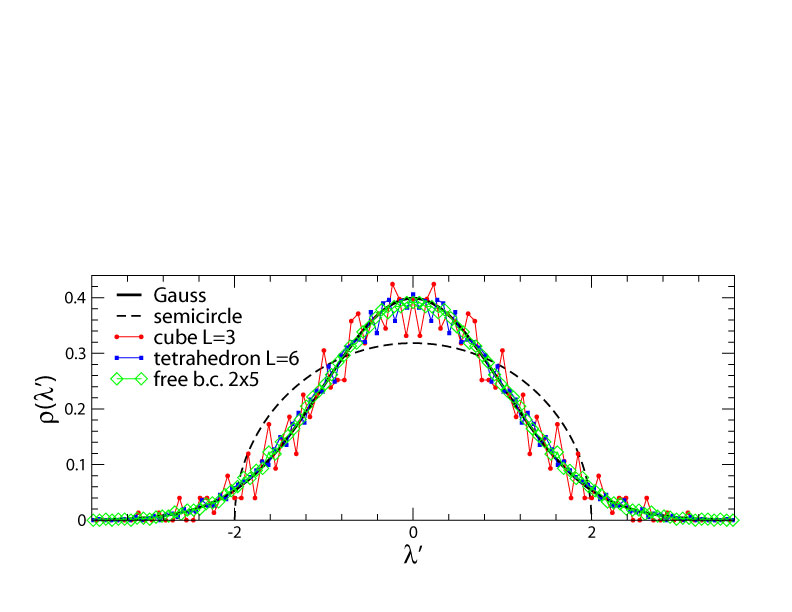}
\caption{\textbf{Spectral densities of phase-space networks.}
Variances are rescaled to 1 by
$\lambda'=\lambda/\bar{k}^{-\frac{1}{2}}$. Black curve: Gaussian
distribution $e^{-\lambda'^2/2}/\sqrt{2\pi}$. Dashed curve: Wigner's
semicircle law for random networks.
$\rho(\lambda)=\sqrt{4\sigma^2-\lambda^2}/(2\pi \sigma^2)$ if
$|\lambda|<2\sigma$ and zero otherwise. The variance $\sigma^2$ is
also rescaled to 1. Red curve: the spectral density of the 980-node
network of $L=3$ cube stacks. Blue curve: the 7436-node network of
$L=6$ sphere stacks in a tetrahedron, i.e.,  $7\times7$ square ice
under the domain wall boundary condition. Green curve: 7782-node
network of $2\times 5$ square ice under the free boundary condition.
Their Gaussian fits are indistinguishable from the black curve.}
\label{fig:specdensity}
\end{figure}

Spectral analysis provides global measures of network properties.
For an $N_n$-node network, the connectivity (or adjacent) matrix
$\mathbf{A}$ is an $N_n\times N_n$ matrix with $A_{ij}=1$ if nodes
$i$ and $j$ are connected, and zero otherwise. Since edges in
phase-space networks are undirected, $\mathbf{A}$ is symmetric and
all its eigenvalues, $\lambda_i$, are real. The spectral density of
the network is the probability distribution of these $N_n$
eigenvalues:
$\rho(\lambda)=\frac{1}{N_n}\sum_{i=1}^{N_n}\delta(\lambda-\lambda_i)$.
$\rho(\lambda)$'s $q$th moment, $M_q$, is directly related to the
network's topological feature.
$D_q=N_nM_q=\sum_{i=1}^{N}(\lambda_i)^q$ is the number of paths (or
loops) that return back to the original node after $q$ steps
\cite{Costa07}. In a bipartite network, all closed paths have even
steps so that all odd moments are zero. Consequently, the spectral
density is symmetric and centered at zero. The $i$th node with $k_i$
neighbours has $k_i$ ways to return back after two steps; hence, the
variance $\sigma^2=M_2=\sum_i k_i/N_n=\bar{k}$, where
$\bar{k}=2N_{edge}/N_n$ is the mean connectivity. We rescale the
spectral densities by $\bar{k}^{1/2}$ to the unit variance (see Fig
\ref{fig:specdensity}). The rescaled spectral densities of different
frustration models collapse onto the same \textit{Gaussian}
distribution. By counting $D_q$, we show that spectral densities are
Gaussian at the infinite-sized limit (see Section I of Supplementary
Information (SI)). This distinguishes phase spaces from other
complex networks. For example, the spectral density of a random
network is the semicircle in Fig.~\ref{fig:specdensity}. The
spectral densities have triangular distributions for scale-free
networks and irregular distributions for small-world, modular
hierarchical and many real-world networks
\cite{deAguiar05,Farkas01}.

Spectral analysis can also detect the network's community (or
modular) structures \cite{Newman06PNAS} if there are any. The
algorithm in ref. \cite{Newman06PNAS} identifies some relatively
highly connected subnetworks (i.e., communities). However, we still
observe a number of edges between subnetworks such that the whole
phase space has to be considered as fully ergodic. Our simulation
shows that the system can easily travel through the whole
phase-space network via basic flips and will not be trapped in a
local community for a long time.

\section{4. Poisson processes and equal probability in phase spaces.}

The fundamental assumption of statistical mechanics is that the
dynamic trajectory of a system wanders through all its phase spaces
and spends the same amount of time in each equally sized region of
the phase space. However, this `equal a priori probability
postulate' (essentially the same as the `ergordic hypothesis'
\cite{Patrascioiu87}) is not necessarily true, as Einstein noted
\cite{Cohen07}. How the system moves from one configuration to the
next depends on the details of the molecules' interactions (e.g.
nearest-neighbor antiferromagnetic interactions here); these
microscopic dynamics may make some configurations more likely than
others.

Network analysis provides an opportunity to study ergodicity. Unlike
billiards with deterministic trajectory, we assume the spin flipping
is due to the random thermal motion and not depends on history. Thus
the dynamical evolution of the system can be viewed as a random walk
on its phase-space network.  It still interesting to ask whether
this random walk can uniformly visit each node given the complex
topology of the network. In another word, whether the system can
visit each possible microstate configuration under the complex
constraint of local nearest-neighbor interactions.

Random walks on a network are rather chaotic, and nodes with higher
connectivities will be visited more frequently. Thanks to the
theorem in ref.\cite{Noh04}, the mean visiting frequency for node
$i$ is $k_i/N_{edge}$, which only depends on local connectivity,
$k_i$, and does not depend on the global structure of the network.
Here, $N_{edge}$ is the total number of edges. This theorem is a
direct consequence of the undirectedness of edges. Although highly
connected nodes are visited more frequently ($\sim k_i$),
interestingly, the equal-probability postulate does not break down
because the average time stayed at node $i$ is $\sim 1/k_i$. Basic
flips are random and independent of history, meaning that it is a
Poisson process. We define the flipping probability of a basic flip
within a unit of time as $\nu$, which is the intensity of the
Poisson process. In Poisson processes, the time interval between
flips (i.e., the staying time) has an exponential distribution,
$e^{-\nu t}$, and the mean staying time is $1/\nu$. Note that
multiple flips will \textit{not} flip \textit{exactly
simultaneously} because time is continuous. Therefore we do not need
to worry about possible illegal configurations if \textit{neighbor}
free spins flip simultaneously. For a node with connectivity $k$,
the superposition of $k$ Poisson processes is still a Poisson
process with intensity $k\nu$ and, consequently, the mean staying
time is $1/(k\nu)$. A random walker has higher frequency ($\sim k$)
to visit a high-$k$ node, but will stay there for a shorter time
($\sim 1/k$), so that the equal-probability postulate is recovered.
Boltzmann assumed that molecules shift from one microscopic
configuration to the next in such a way that every possible
arrangement is equally likely, i.e., all edges have the same weight.
We find that the equal-probability postulate still holds if edges
have different weights (see Appendix B), which, for example, can
represent different potential barriers in complex energy landscapes
in phase spaces.

\section{5. Square ice.} We further study another frustration model
called square ice to identify the more general properties of phase
spaces. Square ice is the two-dimensional version of water ice as
shown in Fig.~\ref{fig:water}. It can be viewed as jigsaw tiling
\cite{Bressoud99} or spin ice
\cite{Bramwel01,Bressoud99,Wang06,Propp01} (see
Figs.~\ref{fig:squareiceall}A,D). Oxygen atoms are represented by
vertices and the relative directions of hydrogen atoms are
represented by arrows. The ground state of the system follows the
\textit{ice rule}, i.e., each vertex has two incoming and two
outgoing arrows. It is also known as the six-vertex model since each
vertex has six possible configurations (i.e., six types of jigsaw
tiles). For a vertex associated with four ferromagnetic spins,
frustration is inevitable (see the example in
Fig.~\ref{fig:squareiceall}C).

Flipping a closed loop of arrows from clockwise to counterclockwise
(or vice versa) does not break the ice rule. The smallest four-spin
loops in Figs.~\ref{fig:squareiceall}D,G are labeled in red
(clockwise) and yellow (counterclockwise). They are basic flips
since any configuration change can be decomposed as a sequence of
such flips \cite{Eloranta99}. Similar to cube stacking, all the
legal configurations of square ice are connected via basic flips
\cite{Eloranta99}. Consequently, the phase-space network of square
ice can be constructed.

The square ices in Figs.~\ref{fig:squareiceall}A,D have domain wall
boundary conditions (DWB) as shown by the black arrows in
Fig.~\ref{fig:squareiceall}D. There is a one-to-one correspondance
between jigsaw tiling with DWB and alternating sign matrices (ASM)
\cite{Bressoud99} (see Fig.~\ref{fig:squareiceall}A). ASM are square
matrices with entries 0 or $\pm$1 such that each row and column has
an alternating sequence of +1 and -1 (zeros excluded) starting and
ending with +1. The number of $n\times n$ ASM is \cite{Bressoud99}
\begin{eqnarray}
W&=&\prod_{1\le i \le j \le n}\frac{n+i+j-1}{2i+j-1}
=\prod_{j=0}^{n-1}\frac{(3j+1)!}{(n+j)!} \nonumber \\
&\sim & \left(\frac{27}{16}\right)^{\frac{n^2}{2}} \textrm{when }
n\to \infty, \label{eq:ASM}
\end{eqnarray} i.e., the number of nodes of the
phase-space network of an $n\times n$ square ice with DWB.

\section{6. Mapping square ices to sphere stacks.} 
\begin{figure}
\centering
\includegraphics[width=1\columnwidth]{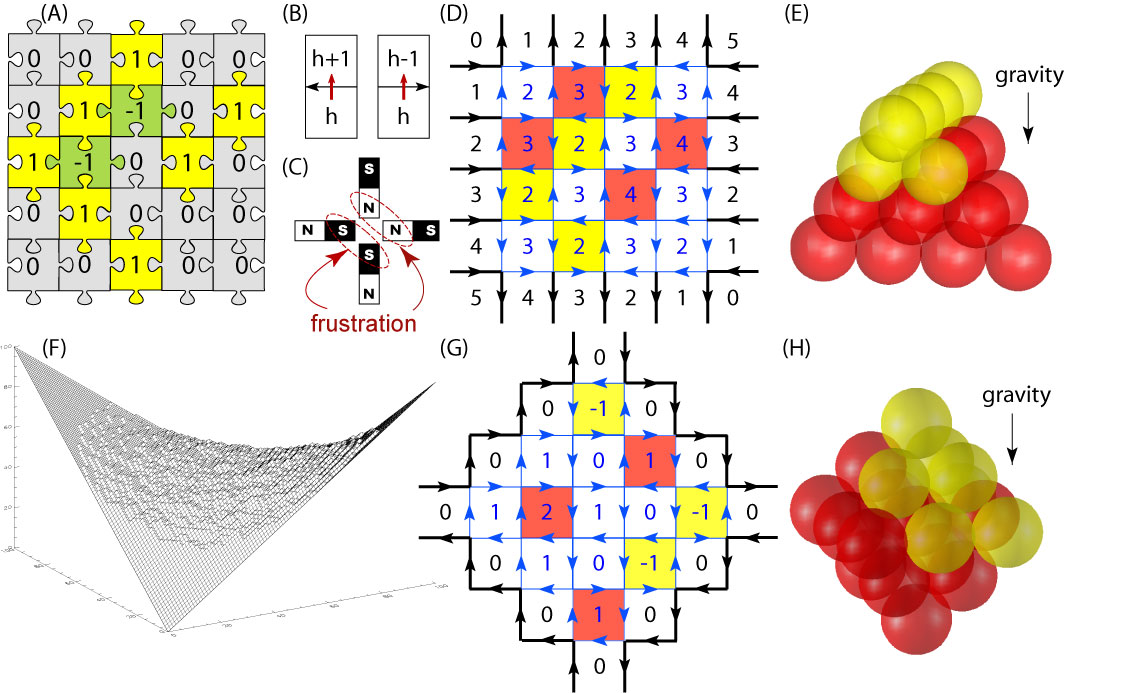}
\caption{\textbf{Square ice.} (A): A $5\times 5$ square ice under
the domain wall boundary condition. Each jigsaw tile can be viewed
as a water molecule with one oxygen atom in the center and two
hydrogen atoms at the two bulges (see Fig.~S4 of SI). By assigning
vertical tiles to be 1, horizontal tiles to be -1 and the other four
types to be 0, a $5\times 5$ alternating sign matrix
\cite{Bressoud99} is obtained. (B): The height rule used in (D) and
(G). (C): Four magnets placed at a cross inevitably have
frustrations. (D): The spin ice mapped from (A). The arrows
represent bulge directions in (A). The blue arrows may flip under
the ice rule. Each plaquette is assigned a height based on the rule
in (B). The upper left corner is defined as height zero. Basic flips
(i.e., four-arrow loops) are labeled in red (clockwise) and yellow
(counterclockwise). (E): The corresponding sphere stack of (D).
Yellow spheres are vacant sites. (F): A typical sphere stack in an
$L=100$ tetrahedron. The sphere centers are connected so that it
appears to be a stack of polyhedra. (G): A spin ice configuration in
an Aztec diamond area under the constant-height boundary condition.
(H): The corresponding sphere stack of (G) in an octahedron.}
\label{fig:squareiceall}
\end{figure}

Mapping 2D triangular
antiferromagnets to 3D cube stacks greatly simplifies the picture of
the phase space and allows combinatorial analysis to generate
quantitative results such as Eq.~\ref{eq:g} and Gaussian spectral
densities. Here, we show that square ices can be mapped to 3D
close-packed spheres in face-centered cubic (FCC) lattices. Each
square plaquette in Figs.~\ref{fig:squareiceall}D,G is assigned a
height \cite{Beijeren77,Eloranta99} based on the rule shown in
Fig.~\ref{fig:squareiceall}B: When walking from the plaquette with
height $h$ to its neighbor, the height increases by 1 if it crosses
a left arrow and decreases by 1 if it crosses a right arrow. The ice
rule guarantees that the height change around a vertex is zero and
$h$ is independent of the path along which it was computed. From the
minimum and maximum possible heights, we found that DWB yields a
stack of building blocks in a tetrahedron (see Fig. S5 of SI). A
plaquette can be flipped only when its four neighbor plaquettes have
the same height. Since each building block is `supported' by four
underneath blocks in an effective `gravity field', the stack can be
viewed as an FCC lattice along the [100] direction, see section III
of SI. Thus, the stacking blocks should be rhombic dodecahedra,
which are primitive unit cells of FCC lattices. FCC lattices can be
conveniently represented by close-packing of spheres. Sphere stacks
in side length $L$ tetrahedra have one-to-one correspondence to
$(L+1)\times (L+1)$ square ices with DWB (see Fig. S5 of SI). The
stack of red spheres in Fig.~\ref{fig:squareiceall}E corresponds to
the configurations in Figs.~\ref{fig:squareiceall}A,D. The physical
heights of the red spheres on the top surface are the heights of the
corresponding plaquettes in the square ice. At the interface between
the red spheres and the yellow vacant sites, the four removable red
spheres on the top surface correspond to the red plaquettes and the
four addable yellow sites correspond to the yellow plaquettes in
Fig.~\ref{fig:squareiceall}C. Similar to the cube stacking, here, a
basic flip from counterclockwise to clockwise (or vice versa)
corresponds to adding (or removing) a sphere. By adding spheres from
the vacant state shown in Figs.~\ref{fig:squareicetetra}A,G, we can
generate all possible stack configurations, i.e., all the nodes of
the phase-space network. Similar to the cube stack case, we can
construct the phase-space network of sphere stacks by adding an edge
between two nodes if the two stacks are different by one sphere.

\section{7. Network properties of sphere stacks.} We numerically studied
phase-space networks of small square ices under various boundary
conditions. Our largest network contains 2068146 nodes and 13640060
edges ($4\times5$ ice under free boundary conditions). All networks
have the small-world property. Their connectivity distributions in
Figs.~\ref{fig:histok}B,C,D and the spectral densities in
Fig.~\ref{fig:specdensity} are similar to those of cube stacks.
Apparently, both cube-stack and sphere-stack phase spaces have
Poisson processes with equal probability and Gaussian spectral
densities.

\section{8. Boundary effects.} Stacks in higher dimensions provide a
vivid means for qualitative visualization of the boundary effect,
which has not been well understood in geometrical frustration
\cite{Millane04}. One peculiar property of geometrical frustration
is that boundary effects often percolate through the entire system
even in the infinite-sized limit \cite{Millane04,Destainville98}.
This can be visualized from a typical sphere stack in the $L=100$
tetrahedron shown in Fig.~\ref{fig:squareiceall}F, which has a
central disordered region and four frozen (ordered) corners known as
the arctic circle phenomenon \cite{Jockush98}. The disordered region
is not uniformly random since different positions have different
mean surface curvatures and entropy densities \cite{Destainville98}
(see Appendix C). Consequently, the infinitely large limit under DWB
\textit{cannot} be called the thermodynamic limit due to the lack of
homogeneity.

Different boundary conditions in square ice correspond to different
container shapes in sphere stacking. For example, the boundary
condition shown in Fig.~\ref{fig:squareiceall}G corresponds to
sphere stacks in an octahedron (see Fig.~\ref{fig:squareiceall}H)
because the lowest possible heights form an inverted pyramid (i.e.
the container) and the highest heights form an upright pyramid (i.e.
the lid). Appendix D shows another boundary condition whose
container and lid have different shapes. At a given boundary
condition, the lid and the container form an interesting pair of
dual surfaces. Some boundary conditions do not have the arctic
circle phenomenon, as illustrated by the sphere stacking in Appendix
C.

\begin{figure}
\centering
\includegraphics[width=0.6\columnwidth]{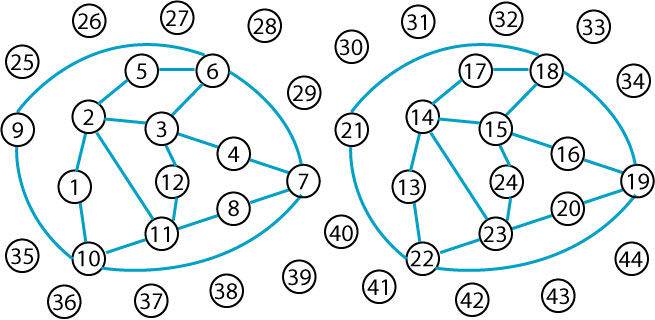}
\caption{\textbf{The phase-space network of $2\times 3$ square ice
under periodic boundary conditions.} There are 44 possible states
(see their detailed configurations in Figs. S9A,B of SI). States 1
to 12 are connected via basic flips; states 13 to 24 are connected;
and states 25 to 44 do not contain any basic flips, thus they are
isolated nodes.} \label{fig:period34phase}
\end{figure}

We found that phase spaces are ergodic under free or fixed boundary
conditions, but nonergrodic under periodic boundary conditions whose
networks consist of disconnected subnetworks. As an example,
Fig.~\ref{fig:period34phase} is the phase-space network of the
$2\times 3$ square ice wrapped on toroid, i.e., under periodic
boundary conditions. It contains two nontrivial (12-node)
subnetworks and 20 trivial isolated nodes. The corresponding 44
configurations are shown in Fig.~\ref{fig:period34all}. For $m\times
n$ periodic square ice wrapped on a toroid, we show that its phase
space contains $2^{n+1}+2^{m+1}-4$ trivial isolated nodes,
$(m-1)\times (n-1)$ nontrivial subnetworks and the smallest
non-trivial subnetwork has $\frac{(m+n-1)!}{(n-1)!(m-1)!}$ nodes
(see Appendix E). These results are confirmed numerically.

\section{9. Discussion and outlook.} We build novel connections between
geometrical frustration, combinatorics (e.g., plane partition and
sphere stacking) and complex networks to exploit open questions and
analysis tools from these fields. Other frustration models, such as
triangular and kagom\'{e} ices, antiferromagnets in 2D kagom\'{e}
and 3D pyrochlore lattices \cite{Moessner98,Lee02}, have height
functions and basic flips so that their phase-space networks can be
similarly constructed. In principle, these models can be mapped to
polyhedra stacking in higher dimensions, so that their rich
symmetries and boundary effects become more transparent.

Quasicrystals can also be mapped to higher-dimensional lattices.
Projecting the high-dimensional lattices to lower dimensions could
result in periodic lattices (i.e., geometrical frustration) at
certain projection angles, or aperiodic structures (i.e.,
quasicrystals) at other angles. Phasons in quasicrystals correspond
to basic flips in geometrical frustration \cite{Destainville98},
thus similar phase-space analysis may be applied to quasicrystals.
In fact, the infinitely degenerated ($\sim e^N$ where $N\to \infty$)
ground states in both geometrical frustration and quasicrystals are
essentially metastable states since the third law of thermodynamics
dictates that the true ground state of real materials must have
finite degeneracy. Network analysis may provide a possible approach
to understanding the observed glassy dynamics in frustrated systems
\cite{Han08}. At finite temperatures, phase-space networks can be
similarly constructed. The nodes are all configurations on the
hypersurface in the phase space determined by the conservation laws.
Configurations change via basic flips and diffusion of thermal
excitations \cite{Han08,Blunt08}. These motions are represented by
edges. The weight of each edge can be assigned by a Boltzmann factor
or defined by the physical details of the real system
\cite{Blunt08}. Height representation can be recovered by assigning
vector heights \cite{Moore00}, so that systems at finite
temperatures might be mapped to stacks in even higher dimensions.

Compared with intensively studied social networks, information
networks, biological networks and technological networks
\cite{Newman03}, phase-space networks belong to a new class with
unique Gaussian spectral densities. A large tool box \cite{Costa07}
has been developed in the recent decade to study network dynamics,
correlations, centrality, community structures, fractal properties
\cite{Song05}, coarse graining \cite{Gfeller07}, etc. These tools
can be readily applied to phase-space studies. In particular, phase
spaces might have fractal structures because stacks of cubes or
spheres have self-repeating patterns on various length scales. This
may cast new light on the highly controversial Tsallis's
nonextensive entropy \cite{Cho02,GellMann04}, which is based on the
assumption that nonequilibrium systems have fractal phase spaces. To
date, a real example to support this assumption has not been
available. Indeed, geometrical frustrated ground states share the
same features as the long-range interacting systems typically
discussed in the context of Tsallis entropy. One example is boundary
effects percolating through the entire system so that the system is
not uniform at the infinite-sized limit and cannot be viewed as a
simple sum of its subsystems (i.e., non-extensive).

In statistical physics, the two models we studied here are
considered as exactly solvable \cite{Baxter82} under periodic
boundary conditions at the infinite-sized limit. Combinatoric
analysis, although challenging, provides an alternative approach to
yield exact results about finite systems and at various boundary
conditions. Cube stacking (i.e., rhombus tiling or plane partition
\cite{Bressoud99}), naturally appears in many chemical and physical
problems, such as counting benzenoid hydrocarbons, percolation,
crystal melting and string theory \cite{Okounkov03}. In contrast to
the intensively studied cube stacking, sphere stacking has not been
explored. Only some combinatoric properties of sphere stacking in
tetrahedra are available since we can map them to ASM. Our numerical
calculations show that there are $2, 7, 42, 429, 7436, 218348\cdots$
ways to pack spheres in $L=1, 2, 3, 4, 5, 6,\cdots$ tetrahedra; and
$2, 18, 868, 230274,\cdots$ ways in $L=1, 2, 3, 4,\cdots$ octahedra.
The former number sequence (i.e., sequence A005130 in
ref.\cite{sequence}) is given by Eq. \ref{eq:ASM}, while the formula
for the latter is not available. Moreover, many questions studied in
cube stacking can be asked about FCC sphere stacking. For example,
how many ways are there to pack $N$ spheres into a tetrahedron? Is
there a similar generating function as cube stacking for sphere
stacking in a tetrahedron \cite{MacMahon}? What is the
ensemble-averaged surface in Fig.~\ref{fig:meansurf100}A, i.e., what
are the entropy density distributions at the infinite-sized limit
\cite{Destainville98,Kenyon07}? These questions can also be asked
about other container shapes. Furthermore, square ice has one-to-one
mappings to other 2D models, such as three-color graphs, dimers,
fully packed loops, etc. \cite{Propp01}. It also has one-to-multiple
mapping to the domino tiling \cite{Elkies92}. Sphere stacking
provides a simple 3D picture and casts new light on these 2D models.

\section{Acknowledgements} We thank Michael Wong for helpful
discussions.

\section{Appendix A: Proof of the Gaussian spectral density}
The characteristic function, i.e., the Fourier transform of the
probability function, uniquely describes a statistical distribution.
It can be written as a series of moments of the distribution. Hence,
to prove that the spectral density is Gaussian, we only need to show
that all orders of the moments are the same as those of a Gaussian
distribution. For a Gaussian distribution centered at 0, its odd
moments are zero and its even moments (of order $q$) are
$M(q)=\frac{(q)!}{2^{q/2}(q/2)!}\sigma^{q}=(q-1)!!\sigma^{q}$, where
$\sigma^2$ is the variance. For an $N_n$-node undirected network,
$D(q)=N_nM(q)$ is the number of directed paths that return to their
starting node after $q$ steps \cite{Costa07}. We count $D_q$ by
stacking cubes/spheres and show that $M(q)=D(q)/N_n$ follows the
Gaussian $M(q)$.

\begin{figure}
\centering
\includegraphics[width=1\columnwidth]{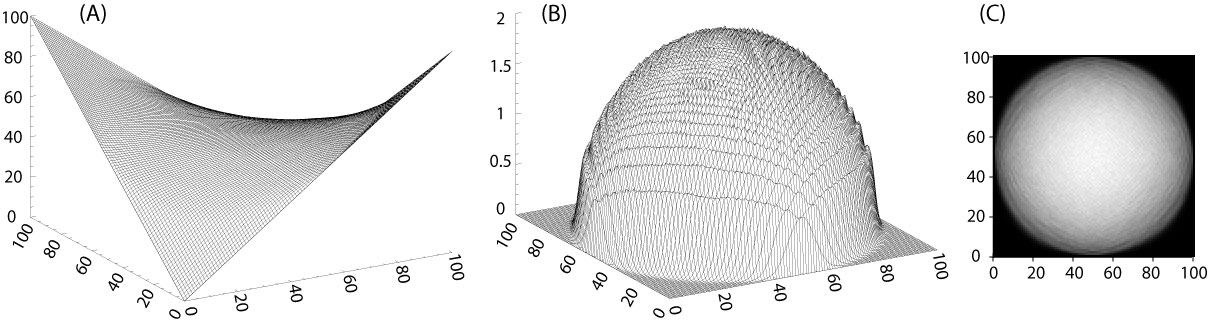}
\caption{(A): The mean sphere stack surface in an $L=100$
tetrahedron. The surface is obtained by averaging over $10^9$ stacks
at equilibrium. A typical surface at equilibrium is shown in Fig. 4F
of the main text. (B): The corresponding probability of basic flips
measured from $10^9$ step simulation. The probability distribution
appears to be a hemisphere. (C): The flipping probability in (B)
represented by brightness. The bright non-frozen region is circular,
which agrees with the arctic circle theorem \cite{Jockush98}.}
\label{fig:meansurf100}
\end{figure}

\begin{figure}
\centering
\includegraphics[width=0.5\columnwidth]{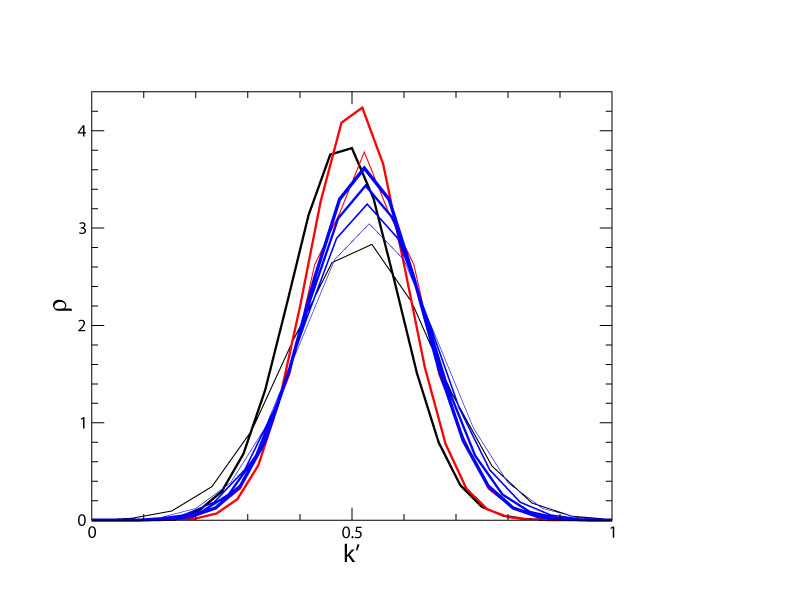}
\caption{The probability distribution of normalized connectivity,
$k'=k/k_{max}$, where the maximum connectivity is $k_{max} \propto
N_{spin} \propto L^2$. This figure is normalized from Fig.~2 in the
main text. Black curves: cube stacks in $L=4$ (thick curve) and
$L=3$ (thin) boxes. Red curves: sphere stacks in $L=7$ (thick) and
$L=6$ (thin) tetrahedra. Blue curves: 2D square stacks (i.e., 2D
sphere stack as shown in Fig.~\ref{fig:2Dstack}A) with $L=11, 10, 9,
8$ (thicker curves for larger $L$). The peaks are in the middle bin
at 0.5, i.e., $k=k_{max}/2$ has the highest number of stack
configurations. In the infinite-sized limit, the normalized
distributions will be asymptotic to a specific functional form. For
a histogram of Fig.~2 in the main text before the normalization, the
peak height, $H$, increases exponentially with the system size,
while the heights at $k_{max}$ are always 1 or 2 (for example, see
Fig.~\ref{fig:cube4}). Thus, their ratio $2/H \to 0$ in infinitely
large systems. This indicates that the asymptotic distribution is
indeed a Dirac delta function as described in ref.
\cite{Destainville98}.} \label{fig:normhistok}
\end{figure}

The phase-space networks are bipartite since walking an odd number
of steps (i.e., adding/removing cubes/spheres an odd number of
times) cannot return back to the original state. Consequently, all
odd moments are zero, i.e., the distribution is symmetric and
centered at 0. $D(2)$ is the number of ways to have one basic flip,
$f_1$, and its reverse flip, $\bar{f}_1$. Given a stack
configuration, $i$, with $k_i$ available basic flips, i.e., node $i$
with connectivity $k_i$ in the phase-space network, its
$D(2)_i=k_i$. Thus, the total $D(2)=\sum_i k_i=N_n\bar{k}$ where
$\bar{k}=2N_{edge}/N_n$ is the mean connectivity. Compared with the
second moment, $M(2)=\sigma^2$, we have $\bar{k}=\sigma^2$. $D(4)$
is the number of ways to have two basic flips, $f_1$, $f_2$, and
reverse flips, $\bar{f}_1$, $\bar{f}_2$. Subscripts 1 and 2 denote
the time order. Given $f_1$ and $f_2$, typically there are three
ways to arrange them in legal order:
$f_1$-$\bar{f}_1$-$f_2$-$\bar{f}_2$,
$f_1$-$f_2$-$\bar{f}_1$-$\bar{f}_2$ and
$f_1$-$f_2$-$\bar{f}_2$-$\bar{f}_1$. Note that the reverse flip,
$\bar{f}_j$, must be later than $f_j$. $f_j$ represents either
adding or removing a cube/sphere. If $f_1$ and $f_2$ are flips of
the same spin or neighbor spins, they are not independent so that
only $f_1$-$\bar{f}_1$-$f_2$-$\bar{f}_2$ is legal. However, the
probability of such a case approaches 0 in infinitely large systems
such that we can neglect the `interference' between basic flips in
large systems and assume that all flips are independent. Next, we
consider how many choices of $f_1$ and $f_2$ we have. Given the
initial state, $i$, $f_1$ has $k_i$ choices and $f_2$ has $k_{i1}$
choices. Here, $k_{i1}$ is the connectivity of a node after walking
one step away from state $i$. In large systems, a dominant number of
states have large $k$ diverging with the rough surface area, $\sim
L^2$. Hence, $k_{i1}\simeq k_i$. Moreover, the dominant number of
states is close to the mean surface, such as
Fig.~\ref{fig:meansurf100}A under the domain wall boundary
condition. The surface shape distribution peaks around this maximum
possible surface and becomes like a Dirac delta distribution when
approaching the infinite-sized limit \cite{Destainville98}. The
probability distribution of normalized connectivity approaches a
Dirac delta distribution as well (see Fig.~\ref{fig:normhistok} and
its caption). Thus the leading term in $k_ik_{i1}$ is $\bar{k}^2$.
At the infinite-sized limit, there are $k_ik_{i1}\simeq \bar{k}^2$
choices of $f_1$ and $f_2$, and three ways to flip them in different
time orders; therefore, $D(4)=3\bar{k}^2$.

Similarly, we count $D(2n)$ by considering $2n$ flips, $f_1,
\bar{f}_1,f_2,\bar{f}_2,\cdots,f_n,\bar{f}_n$. They are placed in a
$2n$-long sequence in time order. First, $f_1$ must be placed at
step 1. Then, there are $2n-1$ choices for placing $\bar{f}_1$.
Then, $f_2$ must be placed at the earliest available step (i.e.,
step 2 if $\bar{f}_1$ is not occupying that step), Then, $\bar{f}_2$
has $2n-3$ choices. Thus, in total, there are $(2n-1)!!$ legal
sequences. Note that this is accurate because a finite number of
$f_j$'s are diluted enough to be considered as independent in an
infinitely large system. Next, we consider how many choices of
$f_j$'s there are. Given the initial state, $i$, $f_1$ has $k_i$
choices, $f_2$ has $k_{i1}$ choices, ... $f_n$ has $k_{i(n-1)}$
choices. Here, $k_{ij}$ is the connectivity of a node after walking
$j$ steps away from the initial node, $i$. $k_{ij}$ depends on the
pathway of the $j$ steps and is not a constant. In total, there are
$\prod_{j=0}^{n-1} k_{ij}$ choices for $\{f_1,f_2,\cdots,f_n\}$ if
node $i$ is chosen as the starting point. When the system size $L\to
\infty$, there will be $N_n(1-\delta)$ states ($\delta \to 0$) with
large connectivity, $k_i \sim L^2$. In cube stacks, adding one cube
can change $k$ by 3 at most because one cube is supported by three
underlying cubes. If the shortest path between two nodes has $j$
steps, their connectivity difference is $|\delta k_{ij}|\le 3j$. In
sphere stacks, one sphere is supported by four underlying spheres;
thus, $|\delta k_{ij}|\le 4j$. Therefore, when $L\to \infty$, $j \ll
L$ and $\prod_{j=0}^{n-1} k_{ij}=\prod_{j=0}^{n-1} (k_i+\delta
k_{ij}) \simeq k_i^n\simeq \bar{k}^n$ by dropping the high-order
terms. The last step uses the fact that the surface shape
distribution becomes a delta distribution when approaching the
infinite-sized limit.

Combining the above results, $D(2n) \simeq \sum_{i=1}^{N_n} (2n-1)!!
\prod_{j=0}^{n-1} k_{ij}\simeq (2n-1)!! N_n \bar{k}^n$, which
becomes exact at the infinite-sized limit. Since $D(2)=N_n
\bar{k}=N_n \sigma^2$, the $2n$th moment of the eigenvalue
distribution $M(2n)=D(2n)/N_n=(2n-1)!!\sigma^{2n}$ is identical to
the $2n$th moment of a Gaussian distribution. Therefore, spectral
densities of phase-space networks are Gaussian at the infinite-sized
limit. In fact, Fig.~3 in the main text shows that spectral
densities are already very close to the Gaussian distribution when
systems are small ($\sim 10^3$ nodes).
\begin{figure}
\centering
\includegraphics[width=0.6\columnwidth]{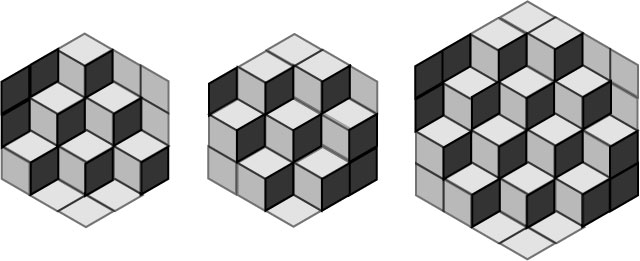}
\caption{The two highest connectivity ($k=13$) states in $L=3$ (odd)
cube stacks and the highest connectivity ($k=24$) state in $L=4$
cube stacks.} \label{fig:cube4}
\end{figure}

\section{Appendix B: Dynamics in phase-space networks with weighted edges}

In the main text, we show that trajectories spend equal amounts of
time at every node on average. This can be easily generalized to
networks with weighted edges, which, for example, can represent
different potential barriers at finite temperatures. We replace an
edge, $i$, with weight $w_i$ with $w_i$ equivalent edges. By
applying such replacement to all edges, we get a new network with
all equally weighted edges. The new connectivity for node $i$ is
$k_i^{w}=\sum_{j=1}^{N_{node}}w_{ij}$. For the same reason as shown
in the main text, the mean staying time at node $i$ is $\propto
1/k_i^w$. The visiting probability is $\propto k_i^w$ by
generalizing the theorem in ref.\cite{Noh04} to weighted networks
\cite{Wu07}. In fact, the proof in ref.\cite{Noh04} can be directly
applied to weighted networks. For a weighted network, the
connectivity matrix, $A_{ij}=w_{ij}$, where $w_{ij}$ is the weight
of the edge between nodes $i$ and $j$. The weighted connectivity is
$k_i^w=\sum_j A_{ij}$. The transition probability from node $i$ to
node $j$ is $A_{ij}/k_i^w$. Suppose a walker starts at node $i$ at
time $t=0$. Then, the master equation for the probability, $P_{ij}$,
to find the walker at node $j$ at time $t$ is given by \cite{Noh04}:
\begin{equation}
P_{i\to j}(t) = \sum_{j_{t-1}} \frac{A_{j_{t-1}j}}{k^w_{j_{t-1}}}
P_{i,j_{t-1}}(t-1). \label{eq:master}
\end{equation}
The transition probability, $P_{ij}(t)$, from node $i$ to node $j$
in $t$ steps can be explicitly expressed by iterating
Eq.~\ref{eq:master},
\begin{equation}
P_{i\to j}(t) = \sum_{j_1,\cdots,j_{t-1}} \frac{A_{ij_1}}{k^w_i}
\frac{A_{j_1j_2}}{k^w_{j_1}} \cdots
\frac{A_{j_{t-1}j}}{k^w_{j_{t-1}}}. \label{eq:itoj}
\end{equation}
By comparing the expressions of $P_{i\to j}(t)$ and $P_{j\to i}(t)$,
we get
\begin{equation}
k_i^w P_{i\to j}(t) = k_j^w P_{j\to i}(t). \label{eq:kp}
\end{equation}
We define the stationary probability, $P_i^{\infty}$, as $t\to
\infty$. Eq.~\ref{eq:kp} implies that $k_i P_{j}^{\infty} = k_j
P_{i}^{\infty}$ and, consequently, we obtain
\begin{equation}
P_i^{\infty} = \frac{k_i^w}{\sum_ik_i^w}=\frac{\sum_i
A_{ij}}{\sum_{i,j}A_{i,j}}. \label{eq:p}
\end{equation}
A random walker visits node $i$ at frequency $P_i^{\infty}\sim
k_i^w$ and stays at node $i$ for the time period $\sim 1/k_i^w$ on
average. Thus the walker spends the same amount of time at each
node.

\section{Appendix C: Square ice and sphere stacks}

\begin{figure}
\centering
\includegraphics[width=0.5\columnwidth]{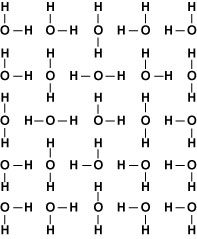}
\caption{(A) Square ice with a domain wall boundary condition. Water
molecules are frozen into a square lattice. This configuration
corresponds to Figs. 4A, D, E in the main text.} \label{fig:water}
\end{figure}
\begin{figure}
\centering
\includegraphics[width=1\columnwidth]{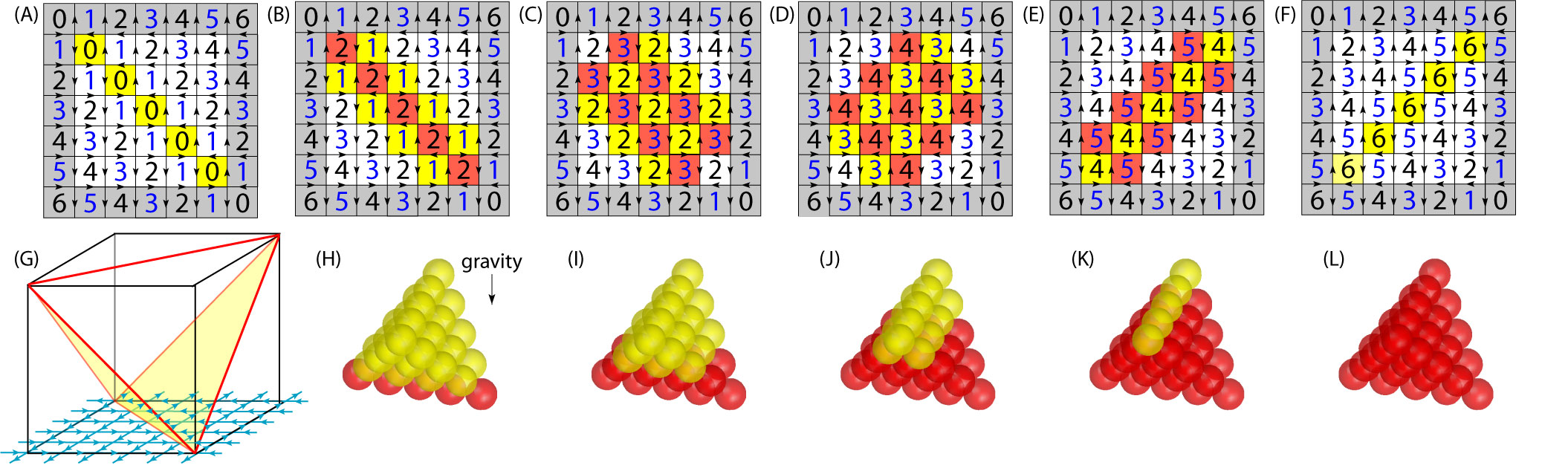}
\caption{The one-to-one mapping between square ice and sphere
stacks. (A-F): $6\times 6$ square ices. The domain wall boundary
condition is labeled in grey. The height of each square plaquette is
labeled with a number ranging from 0 to 6. The upper left corner is
defined as zero height. Other heights are generated by the height
rule in Fig.~5B in the main text. (A): the vacant state with the
lowest possible heights. Heights in (F) define a container
consisting of two yellow triangles shown in (G). Counterclockwise
four-spin loops are labeled in yellow. Flipping all of them in (A)
leads to the configuration in (B). Clockwise four-spin loops are
labeled in red. Flipping the yellow plaquettes results a series of
states shown in (C, D, E, and F). (F) has the highest possible
heights without yellow plaquettes. These heights define a lid, which
is an upside-down container in (G). All legal square ice
configurations can be generated by flipping yellow plaquettes from
the vacant state or, reversely, by flipping red plaquettes from the
full state. A plaquette can be flipped only when its four neighbour
plaquettes have the same height. The height of each plaquette is the
physical height of the corresponding spheres on the top surface of
the stack. Each basic flip can be viewed as adding or removing a
sphere. (H-L): the red sphere stacks corresponding to (B-F). The
yellow spheres are vacant sites.} \label{fig:squareicetetra}
\end{figure}

Figure~\ref{fig:water} shows a $5\times 5$ square ice with the
domain wall boundary condition (DWB) corresponding to Figs. 4A, D,
E. in the main text. Figure~\ref{fig:squareicetetra} shows $6\times
6$ spin ices with DWB and their corresponding FCC sphere stacks in
an $L=5$ tetrahedron. Flipping all counterclockwise four-spin loops
to clockwise corresponds to adding one layer of spheres. The
tetrahedron emerges from the maximum packing, i.e., the full state.

A square plaquette can flip only when its four neighbors have the
same height, which indicates that one building block in the 3D stack
should be supported by four building blocks underneath. Thus, the
stack can also be viewed as a body-centered cubic (BCC) lattice
\cite{Beijeren77}. Both FCC and BCC stacking have the same
combinatoric properties since they are only different by a stretch
(see a 2D analogy in Fig.~\ref{fig:2Dstack}A). FCC is better than
BCC because (1) FCC can be viewed as close-packed spheres in simple
container shapes; (2) FCC has simple polyhedra stacking under
gravity. The Wigner-Seitz cell of an FCC lattice is a rhombic
dodecahedron, which is supported by four rhombic dodecahedra
underneath, while the Wigner-Seitz cell of a BCC lattice can be
supported by one block underneath since it has a flat square on the
top.

\section{Appendix D: Boundary effects}

\begin{figure}
\centering
\includegraphics[width=0.8\columnwidth]{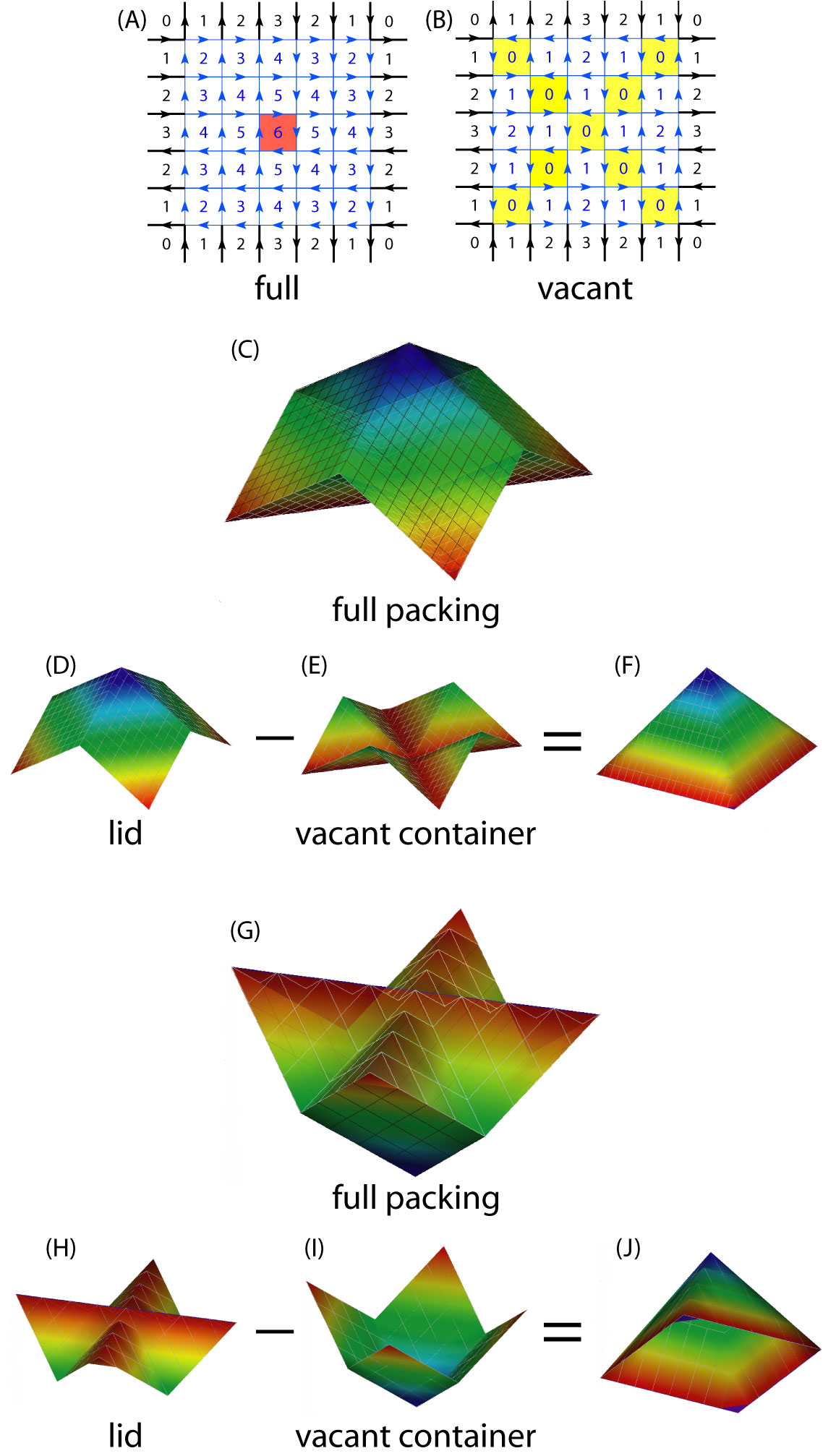}
\caption{$6\times 6$ square ice with the `alternating boundary
condition'. (A): The maximum possible heights, i.e., the lid,
contains one clockwise basic flip labeled in red. (B): the minimum
possible height, i.e., the container, contains multiple
counterclockwise basic flips labeled in yellow. (C): The 3D shape of
a full stack, including a lid (D) and a container (E). (F): The
height difference between the lid and the container is a pyramid.
(G-J): Upside-down geometries of (C-F). The lid (H) emerges from the
maximum packing in the container (I), which is the upside-down
version of (D).} \label{fig:alterboundary}
\end{figure}

We use the `alternating boundary condition' shown in
\ref{fig:alterboundary}A, B to elucidate the lid-container duality.
Given a boundary condition, the minimum (or maximum) possible
heights can be directly written out, for example see
Figs.~\ref{fig:alterboundary}A, B. These heights define the lid and
the container. The lid and the container have different shapes. The
container contains multiple height minima and the lid contains one
height maximum (see the colored squares in
Figs.~\ref{fig:alterboundary}A, B). They form an interesting pair of
dual surfaces: using one as the container, the other will emerge as
the surface of the highest `sand pile' of small spheres. In fact,
given a fixed boundary condition, the container and the lid are dual
surfaces because, if we reverse the height rule in Fig.4(B), the
container and the lid switch roles. Packing spheres in the container
is equivalent to packing buoyant spheres in the corresponding lid.
The height difference between a lid and container is a pyramid; for
example, see Figs.~\ref{fig:alterboundary} and
\ref{fig:tetracontainer}.

\begin{figure}
\centering
\includegraphics[width=1\columnwidth]{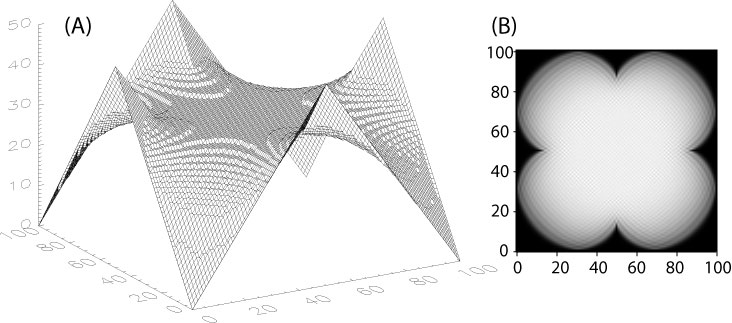}
\caption{(A): The ensemble-averaged height surface for $L=100$
square ice with alternating boundary condition shown in
Fig.~\ref{fig:alterboundary}A, B. The heights are rounded off to
integers to show equal-height contours. The container shape is shown
in Fig.~\ref{fig:alterboundary}E. (B): The flipping probability
represented by the brightness.} \label{fig:tilt100meanandfluc}
\end{figure}

\begin{figure}
\centering
\includegraphics[width=0.8\columnwidth]{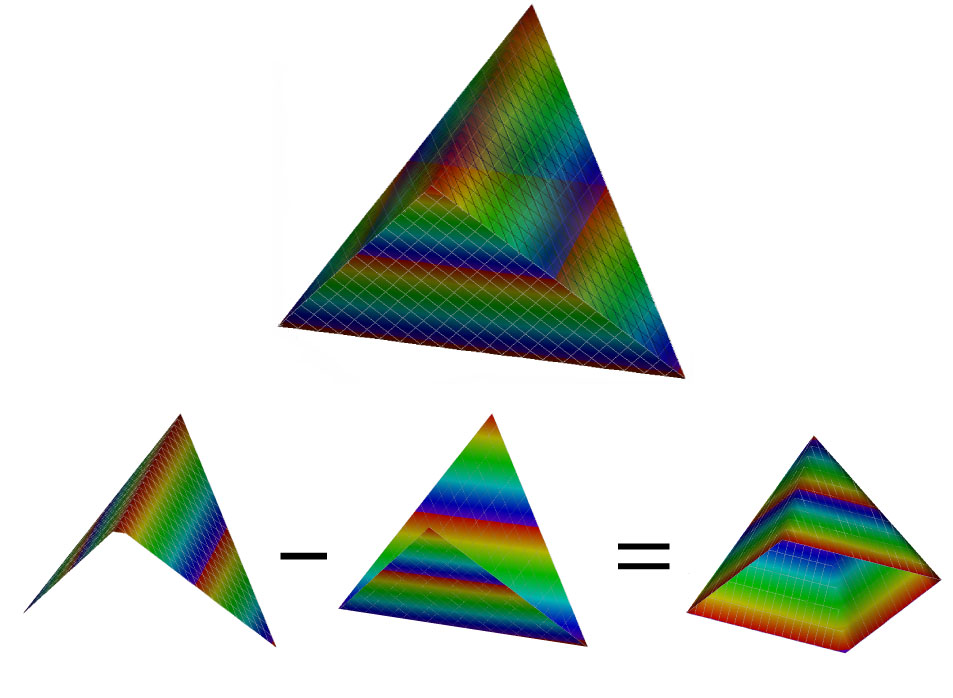}
\caption{The lid, container and their height difference of a
tetrahedron.} \label{fig:tetracontainer}
\end{figure}

Some boundary conditions do not exhibit the arctic circle phenomenon
shown in Fig.~\ref{fig:meansurf100}. Their disordered region may not
have a circular shape or may not even have a frozen area
\cite{Eloranta99}. The container shape provides an intuitive
understanding about how boundaries affect the disordered region. In
a tetrahedron, the largest horizontal cross section is a square in
the middle height that is circumscribed by the disordered region. In
an octahedron, the largest horizontal cross section is the total
square ice area so that there is no frozen region under the boundary
condition shown in Fig.~4G of the main text. This is confirmed by
our simulation. Other boundary conditions may lead to the
non-circular disordered region. For example, the flower shape
observed in ref. \cite{Eloranta99} (see
Fig.~\ref{fig:tilt100meanandfluc}B) is a direct consequence of the
container shown in Fig.~\ref{fig:tilt100meanandfluc}A.

The ensemble average over random stacks results a mean surface shown
in Fig.~S1A. At the infinite-sized limit, dominate states are close
to this mean surface \cite{Destainville98}, i.e., the typical
surface in Fig.~4F approaches the mean surface in Fig.~S1A when
$L\to \infty$. In the space of the stack surface, the distribution
peaks around this maximum possible surface and becomes more and more
like a Dirac delta distribution when approaching the infinite-sized
limit \cite{Destainville98}. The local gradient of the surface
determines the density of the basic flips, i.e., the density of
configurational entropy $s_0$ \cite{Destainville98}. In Fig.~S1A ,
the $s_0$ is zero in the frozen areas and continuously varies to
reach its maximum value near the center of the square ice, with a
non-zero gradient everywhere except near the center
\cite{Destainville98}. Consequently, the infinitely large limit of
the DWB \textit{cannot} be called the thermodynamic limit due to the
lack of homogeneity. In contrast, the boundary condition in Fig.~4G
has the thermodynamic limit since the limiting surface in the
octahedron is flat everywhere. The flat surface has the maximum
possible $s_0$ (for example, see Fig.~S3) so that its spatial
averaged, $\bar{s}_0$, is as high as that of the free boundary
condition. The boundary condition in Fig.~4G is a subset of the
periodic boundary condition, so that the periodic boundary condition
has the same $\bar{s}_0$ as the free boundary condition at the
infinite-sized limit. This explains why the $\bar{s}_0$ calculated
from the periodic boundary condition \cite{Nagle66} agrees so well
with the experimental results on water ice obtained under the free
boundary condition \cite{Giauque33}. When the height difference of a
fixed boundary is comparable to $L$, the limiting surface is not
flat and $\bar{s}_0$ is smaller. For example, the zero-point entropy
of $L\times L$ square ice in the infinite-sized limit under DWB is
$\bar{s}_0=k_BL^2\ln{N_n}= k_B \ln(\sqrt{27/16})$ \cite{Elkies92}
based on Eq.~2 in the main text, which is smaller than $k_B
L^2\ln(\sqrt{64/27})$ under the periodic boundary condition
\cite{Lieb67}. For cube stacks, $\bar{s}_0=k_B \ln(\sqrt{27/16})$
under the hexagonal boundary condition based on Eq.~1 in the main
text, which is smaller than $0.323k_B$ under the periodic boundary
condition \cite{Wannier50}.

The typical stack configuration in a tetrahedron or octahedron is
$\sim$ 50\% filled because the lid and the container have the same
shape. When they have different shapes, a typical stack
configuration may not be $\sim$ 50\% filled. Figure
\ref{fig:tilt100meanandfluc}A is the averaged surface in an $L=100$
container. The total volume has $L(L^2-1)/6$ spheres, and $\sim$42\%
of the volume is filled with spheres on average. Note that the
largest horizontal cross-section is at $h=\sqrt{2}/3$ with a
corresponding filled fraction of 4/9.

\section{Appendix E: square ice with periodic boundary condition}
\begin{figure}[!t]
\centering
\includegraphics[width=1\columnwidth]{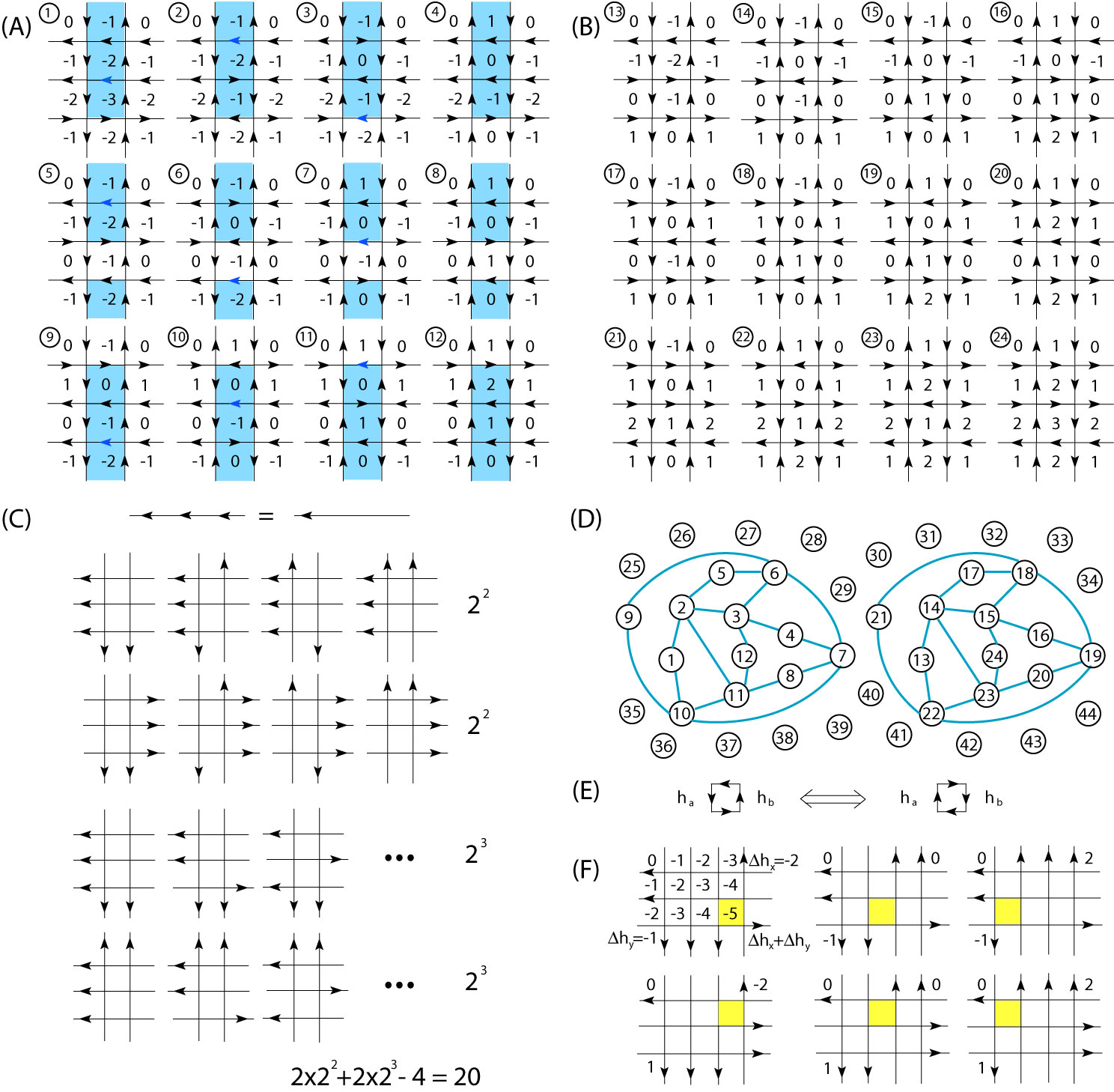}
\caption{Configurations and the phase-space network of $2\times 3$
square ice wrapped on a toroid, i.e., with the periodic boundary
condition. (A): Configurations 1 to 12. The upper left plaquette is
defined as zero height and other heights are derived from it by
using the height rule in Fig.~4B in the main text. (B):
Configurations 13 to 24. (C): 20 configurations that contain no
basic flip. They are categorized into four types: all horizontal
spins are (1) leftwards; (2) rightwards; all vertical spins are (3)
upwards; (4) downwards. Four configurations are double counted, so
the total number of configurations is $2\times 2^2+2\times 2^3-4$.
(D): The phase-space network. Two nodes are connected if they are
different by one basic flip (i.e., the flip of a four-spin loop).
(E): The flip of a spin loop does not change the height difference,
$h_a-h_b$ if $h_a$ and $h_b$ are both inside or outside the loop.
(F): $m\times n$ square ice ($m=3$, $n=4$) with periodic boundary
conditions. There are $(m-1)\times (n-1)=6$ types of different
$(\Delta h_x,\Delta h_y)$. Configurations with the same $(\Delta
h_x,\Delta h_y)$ form one cluster by adding/removing spheres in the
corresponding container, while configurations with different
$(\Delta h_x,\Delta h_y)$ are disconnected because zero energy flips
cannot change the height mismatches. Each state has a unique lowest
height plaquette labeled in yellow.} \label{fig:period34all}
\end{figure}

Here, we use the $2\times 3$ square ices to illustrate that periodic
boundary conditions result in disconnected phase-space networks.
Figure~\ref{fig:period34all} shows the 44 configurations of the
$2\times 3$ square ice wrapped on a toroid. Note that this periodic
boundary condition is for spins, not for heights. The upper left
corner is defined as zero height. The 12 configurations in
Fig.~\ref{fig:period34all}A are connected by basic flips and form a
12-node cluster as shown in Fig.~\ref{fig:period34all}D. The other
12 configurations in Fig.~\ref{fig:period34all}B form another
12-node cluster in Fig.~\ref{fig:period34all}D. The height
difference between the top corners and bottom corners is +1 in
Fig.~\ref{fig:period34all}A and -1 in Fig.~\ref{fig:period34all}B.
Note that the four corners are essentially the same plaquette on the
toroid, so they must be either all inside or all outside of a loop,
such as the one shown in Fig.~\ref{fig:period34all}E. Consequently,
the height difference between the corners cannot be changed by
flipping a closed spin loop. Therefore, the nodes in
Fig.~\ref{fig:period34all}A and B form two disconnected clusters.
The 20 isolated nodes in Fig.~\ref{fig:period34all}D correspond to
the 20 configurations in Fig.~\ref{fig:period34all}C, which contain
no basic flips.

We generalize the above results to the $m\times n$ periodic lattice
and prove that it has $(m-1)\times (n-1)$ non-trivial clusters and
$2^{n+1}+2^{m+1}-4$ isolated nodes. Unlike fixed boundary
conditions, after walking along a closed loop in the $x$ or $y$
direction on a toroid and coming back to the original plaquette, the
height may change. Such height differences, $\Delta h_x$ and $\Delta
h_y$, uniquely characterize each disconnected subnetwork. Consider
an arbitrary plaquette on a toroid. We unwrap the lattice onto a
plane so that this plaquette is at the upper left corner with the
height defined as 0, e.g., see Fig.~\ref{fig:period34all}F. As shown
in Fig.~\ref{fig:period34all}E, any zero energy flip of a spin loop
cannot change the height differences between the four corners since
they are essentially the same plaquette on the toroid. Consequently,
configurations with different $\Delta h_x$ or $\Delta h_y$ cannot be
connected by basic flips. On the other hand, if configurations have
the same $\Delta h_x$ and $\Delta h_y$, they must be connected
because they have essentially the same fixed boundary condition
after being unwrapped onto a plane (see
Fig.~\ref{fig:period34all}F). Here, we use the fact that all legal
configurations at a fixed boundary condition are connected by basic
flips \cite{Eloranta99}. Since they are connected, we can choose the
configuration whose bulk spins are along the boundary spins to
represent each subnetwork (see examples in
Figs.~\ref{fig:period34all}C, F). Next, we consider the number of
representative configurations, i.e., the number of subnetworks. If a
representative configuration has no basic flips, all of its
horizontal spins or vertical spins have to be along the same
direction as shown in Fig.~\ref{fig:period34all}C. The `energy
barriers' between these states are large for large systems because
$m$ or $n$ spins need to flipped simultaneously in order to change
from one state to another without breaking the ice rule. If all
horizontal spins are leftwards (or rightwards), there are $2^m$
configurations for vertical spins (see Fig.~\ref{fig:period34all}C).
If all vertical spins are upwards (or downwards), there are $2^n$
configurations for horizontal spins. In total, the four
configurations are double counted so that there are
$2^{n+1}+2^{m+1}-4$ isolated nodes, i.e., configurations without
basic flips. Next, we consider nontrivial clusters with multiple
nodes. The corner height difference, $\Delta h_x$, has $m-1$
possible values, and $\Delta h_y$ has $n-1$ possible values (see
Fig.~\ref{fig:period34all}F), so that there are $(m-1)\times (n-1)$
nontrivial subnetworks in total. The representative configurations
of the six subnetworks shown in Fig.~\ref{fig:period34all}F are
chosen to have the lowest possible heights, i.e., vacant containers
for sphere stacking. Each configuration is characterized by one
basic flip labeled as yellow squares, i.e., the lowest point of the
vacant container. Apparently, there are $(m-1)\times (n-1)$
positions for a yellow square, i.e., $(m-1)\times (n-1)$
subnetworks. This result confirms that the zero-point entropy
$\bar{s}_0$ of the whole network is the same as that of the largest
subnetwork under the constant-height boundary condition (see section
IV) because $(m-1)\times (n-1)$ is logarithmically small compared
with the total number of configuration $\sim e^{N_{spin}}\sim
e^{m\cdot n}$.

\begin{figure}[!t]
\centering
\includegraphics[width=0.5\columnwidth]{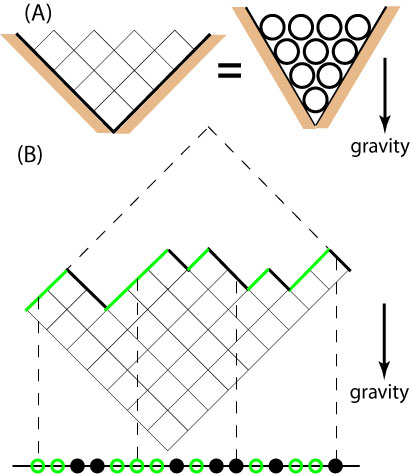}
\caption{(A): 2D circle stacking is combinatorically equivalent to
2D square stacking because each circle or square is supported by two
circles underneath in a gravity field. (B): Mapping a 2D stack of
squares in an $a\times b=7\times 9$ container to a chain of $a$
solid particles and $b$ holes \cite{Okounkov03}. The dynamics, i.e.,
the diffusion of particles, is described as a symmetric simple
exclusion process (SSEP). The number of 2D stack configurations in a
container is $C_{a+b}^a=(a+b)!/(a!b!)$, i.e., the number of ways to
put $a$ particles onto $a+b$ sites.} \label{fig:2Dstack}
\end{figure}

Next, we show that the smallest nontrivial cluster has
$\frac{(m+n-1)!}{(n-1)!(m-1)!}$ nodes. In Fig.
\ref{fig:period34all}F, the two middle configurations represent
132-node subnetworks and the other four configurations represent
60-node subnetworks. When the yellow square is at the corner, the
height function indicates that the container shape is a tilted 2D
container rather than a 3D container. Thus, the number of spheres
packing in such a container is much smaller than that in 3D
containers whose lowest point (the yellow plaquette) is not at the
corner. To count the number of states in a tilted 2D container, we
first consider the simple case in Fig.~\ref{fig:period34all}A.
Configuration 1 in Fig.~\ref{fig:period34all}A is the representative
state with the lowest height of -3. Configurations 1 to 4 in
Fig.~\ref{fig:period34all}A have the same boundary spins so that we
can view them as the 2D sphere stacks in the same $1\times 3$-sized
2D rectangle. Such a blue $1\times 3$ container has three possible
positions relative to the zero height plaquette (see configurations
1, 5 and 9 in Fig. \ref{fig:period34all}A). In total the subnetwork
has $3\times 4=12$ nodes. It is easy to generalize this counting to
$m\times n$ square ice on a toroid. There are $m$ possible positions
for the $m\times (n-1)$-sized rectangle. With 2D sphere stacking in
an $a\times b$ container, there are $C_{a+b}^{a}=(a+b)!/(a!b!)$
configurations (see Fig.~\ref{fig:2Dstack} and its caption).
Consequently, there are
$mC_{m+n-1}^{m}=\frac{(m+n-1)!}{(n-1)!(m-1)!}$ nodes in the smallest
nontrivial subnetworks. We confirm the above results numerically.
Our numerical results also confirm the number sequence A054759 in
ref.\cite{sequence} for the $n\times n$ square ice under periodic
boundary conditions.

\end{document}